
\magnification=\magstep1
\baselineskip=24pt
\nopagenumbers
\parindent=0.3truecm
\parskip=3pt
\font\sixteenrm=cmr17 at 16pt

\def\etal{{\it et~al.\ }}

\def\ie{{\it i.e.,~}}

\def\ltsima{$\; \buildrel < \over \sim \;$}
\def\simlt{\lower.5ex\hbox{\ltsima}}
\def\gtsima{$\; \buildrel > \over \sim \;$}
\def\simgt{\lower.5ex\hbox{\gtsima}}
\def\ref{\noindent\hangindent.5in\hangafter=1}

\vskip 8truepc
\centerline{\sixteenrm  RADIO EMITTING DUST IN THE }
\vskip 1truecm
\centerline{\sixteenrm  FREE ELECTRON LAYER OF SPIRAL GALAXIES:}
\vskip 1truecm
\centerline{\sixteenrm  TESTING THE DISK/HALO INTERFACE}
\vskip 5truepc
\centerline{\bf A. Ferrara $^{1,2,3}$ {\rm and} R.-J. Dettmar $^{1,3}$}
\centerline{\it ${^1}$ Space Telescope Science Institute, 3700
San Martin Drive, Baltimore, 21218 MD}
\centerline{\it ${^2}$ Osservatorio Astrofisico di Arcetri,
Largo E. Fermi, 5, 50125 Florence, Italy}
\centerline{\it ${^3}$ Affiliated with the Space Science Department, ESA}
\vskip 7truecm
To be submitted to: {\it The Astrophysical Journal}
\vfill\eject
\centerline{\bf ABSTRACT}
We present a study of the radio emission from rotating, charged dust grains
immersed in the ionized gas constituting the thick, H$\alpha$-emitting disk
of many spiral galaxies. Using up-to-date optical constants, the charge
on the grains exposed to the diffuse galactic UV flux has been calculated.
An analytical approximation for the grain charge has been derived,
which is then used to obtain the grain rotation
frequency. Grains are found to have substantial radio emission peaked
at a cutoff frequency in the range 10-100~GHz, depending on the grain size
distribution and on the efficiency of the radiative damping of the grain
rotation. The  dust radio emission is compared to the free-free
emission from the ionized gas component; some constraints on the
magnetic field strength in the observed dusty filaments are also discussed.
The model can be used to test the disk-halo interface environment in spiral
galaxies, to determine the amount and size distribution of dust in
their ionized component, and to investigate the rotation mechanisms for the
dust. Numerical estimates are given for experimental purposes.
\vfill\eject
\centerline{\sl 1. MOTIVATION}
\medskip
Interstellar dust is known to be responsible for redistributing the  energy in
the interstellar
radiation field mainly by reemitting in the far-infrared  (e.g., Mathis 1990).
In the following it is demonstrated that the classical problem (Chandrasekhar
1943, Erickson 1957,
Hoyle \& Wickramasinghe 1970) of charged and spinning dust
particles caused by interaction of grains with the surrounding gas and
radiation field will lead to
radiation at  easily accessible  radio continuum wavelengths if the
distributions of grain
sizes and charges are properly taken into account. As a consequence, it is well
possible that in specific
astrophysical environments this dust radio emission contributes to the observed
radio continuum
spectra in addition to thermal and non-thermal processes.
\par
Of particular interest in this respect is the thick gas layer of spiral
galaxies.
The large scale-height of the, most probably,
photoionized gas layer in some galaxies such as the Milky Way (Reynolds 1991)
is indicative of
an intensive UV-radiation field high above the plane. In addition,
the presence of dust in these disk/halo interfaces of spiral galaxies on kpc
scales has been
suggested by theoretical (Franco et al. 1991)  and observational work (Keppel
et al. 1991).
This is an interesting situation  to consider  the
scenario developed in this paper. The application to such an environment could
result in
independent determinations of dust masses, grain size distributions,
gas-to-dust ratios, etc.

\bigskip
\centerline{\sl 2. GRAIN CHARGE}
\smallskip
In this Section we will derive the charge, $Z$, on grains located
in the H$\alpha$ emitting gas known to extend to heights well in excess
of the main gaseous disk of the parent spiral galaxy. Grains interact
with the surrounding (partially ionized) gas, with the radiation
field photons and with other grains; we will neglect however the last
possibility given the low collision rate of the process.
Throughout the paper we will adopt the values given by Reynolds (1993)
($T_e=10^4$~K, $n_e=0.1$~cm$^{-3}$) for the electron temperature and density in
our Galaxy, respectively.
\par
The UV radiation field at high galactic latitudes, the one relevant for
the present calculation, has been
subject of several studies in the last years and it is still subject of
debate. A rather complete collection of
observational results has been compiled by Reynolds (1990). The data
are well fitted by a power-law extending approximately from 8 to 300 eV,
with the flux expressed as
$${\cal F}_\nu = {\cal F}_{\nu_L} \epsilon_{Ry}^{-4},\eqno(2.1)$$
where ${\cal F}_{\nu_L}=10^4$~phot~cm$^{-2}$~s$^{-1}$~sr$^{-1}$~eV$^{-1}$
is the flux at the Lyman limit and $\epsilon_{Ry}$ is the photon energy
expressed in $Ry$; eq. (2.1) is also consistent with data not included
in the collection (F. Paresce, private communication). Some uncertainty is
present in the determination of the lower and upper limits to ${\cal F}_\nu$,
but for our aims this does not represent a problem as the lower limit is
dictated by the threshold energy for the photoelectric effect ($B\sim 8$~eV
for most grain materials) and the upper limit, $\epsilon_{max}\gg B$, by the
optical properties of the grains (above 300~eV they
are virtually transparent). An useful parameter to be defined is
$\Xi={\cal F}_{\nu_L}/n_e T_e$ proportional to the ionization parameter.
\par
The second necessary ingredient is the absorption coefficient,
$Q_{abs}(a,\epsilon)$,
where $a$ is the grain size, of the dust in the UV/soft X-rays range. We have
calculated numerically $Q_{abs}(a,\epsilon)$ through the Mie formulae for
spherical silicate grains using the dielectric constants kindly provided by
Martin \& Rouleau (unpublished, but some results are
in Martin \& Rouleau 1991).  These
constants are an extension of the Draine \& Lee (1984) ones for the
astronomical silicate up to 16~eV; from 16~eV to 31~eV they are extrapolated
from Huffman (1975) and above 31~eV a synthesized FeMgSiO$_4$ cross section
has been used. The results are shown in Fig.~1 for different grain sizes.
In the range of the scattering parameter $x=2\pi a/\lambda$ we are interested
in, none of the classical physical approximations (Rayleigh-Gans,
geometric optics) can be used for all the sizes. Since it is useful to have
an analytical approximation anyway, we have fitted the numerical results with
the following law:
$$Q_{abs}(a,\epsilon)=\left(\epsilon\over 8~{\rm eV}\right)^{2.5[(a/0.1\mu
m)^{0.1}-1]}
=\left(\epsilon\over 8~{\rm eV}\right)^{\gamma_a}\eqno(2.2)$$
We will make use of this simple law, which is rather rough for
large grains but it is reasonably accurate for $a\simlt 0.1 \mu$m,
to calculate the average absorption coefficient weighted on the
spectrum: $\langle Q_{abs}(a)\rangle=3/(3-\gamma_a)$.
\par
We have solved the detailed balance equation for the grain charge numerically;
the results are shown in Fig. 2 for different values of ${\cal F}_{\nu_L}$.
The explicit form of such an equation is given in the Appendix; here we just
mention the processes that we have included in the calculation. In a pure
hydrogen
plasma, as the one we have assumed, positive charge is accumulated on the grain
surface by i) proton collisions; ii) secondary electron emission following ion
collisions; iii) photoelectric effect. Negative charging processes are i)
electron collisions ii) secondary electron emission following electron
collisions. Field emission, which limits the amount of charge on the grain,
has been taken into account even if, in the conditions of interest, this
limiting value has never been reached.
Field ionization of the H atoms coming from impinging protons
that have recombined
with $\pi$-electrons of the lattice is negligible for the grain
electrostatic potential of interest here.
Since thermal sputtering yields are negligible below
$T\simlt 10^5$~K, there are two mechanisms that may modify and limit the grain
size distribution. Shattering due to electrostatic stress $S_e$ could destroy
very small grains: however, for the adopted parameters ($\Xi=10$)
and a tensile stress $\tau=10^{10}$~ergs~cm$^{-3}$ (Draine \& Salpeter 1979)
this does not affect grains with
$$a\simgt 2\times 10^{-8} \left({\tau\over 10^{10} {\rm ergs~
cm}^{-3}}\right)^{-1/2} \left({\eta\over 1.49\times 10^7{\rm cgs}}\right)~{\rm
cm}\eqno(2.3)$$
($\eta$ is defined in eq. [2.4]).
The second possible effect is related to the centrifugal stress $S_c$
arising from the grain rotation:
we will discuss this point in the next Section.
\par
The left panel of Fig. 2 shows the grain charge as a function of $a$
for two different values of ${\cal F}_{\nu_L}$; in the right panel the rates
for
the various charging processes are shown for the standard case
${\cal F}_{\nu_L}=10^4$ ($\Xi=10$). Clearly, the stronger the radiation field
the smaller is the value of the negative charge.
\par
Looking at the various rates it is immediately evident that a simple
approximation can be attempted: in this case,
only two processes, \ie ion and electron collisions,
substantially contribute to the charge equilibrium. Thus we can find an
approximate solution of the detailed balance (eq. [A1]) obtained considering
only the two above mentioned processes. A simple linear relation
between $Z$ and $a$ approximates very accurately the numerical
solutions found (see Appendix):
$$Z(a)=-2.504 {kT_e\over e^2} a=-1.49\times 10^7 \left({T_e\over
10^4~{\rm K}}\right)a =-\eta a;\eqno(2.4)$$
for the adopted $T_e$,  $\eta=1.49\times 10^7$ (cgs units).
Eq. (2.3) is also a restatement of the classical result obtained by
Spitzer (1941) and it is indicative of the scarce relevance of the
radiation field in determining the grain charge.

\bigskip
\centerline{\sl 3. DUST RADIO EMISSION }
\smallskip
Grains interact with the gas and the radiation field photons and
they are put into rotation. There are a number of random torques that
need to be considered in order to calculate the steady  rms angular velocity
$\omega$ of grains: collisions with gas atoms, absorbed or emitted
photons, photoelectric emission and molecule formation on the grain surface.
The various impulsive torques are supposed to act randomly in direction, thus
the angular momentum grows in a random walk; if surface irregularities are
present torques may be not time-averaged and therefore lead to a suprathermal
rotation. In order to evaluate $\omega$ we will rely on the results obtained
by Purcell (1979).
\par
The rms rotational velocity $\langle \omega \rangle$ can be obtained in a
straightforward manner once the effective temperature $T_{eff}$ characterizing
the Brownian motion is known from
$${1\over 2} I \omega ^2 = {3\over 2}k T_{eff},\eqno(3.1)$$
where $I$ is the rotational inertia.
If the only processes exciting rotation would be
atom-grain collisions and atom evaporation from the grain surface then
$T_{eff}$ would be exactly the arithmetic mean of $T_e$ and $T_d$,
where $T_d$ is the dust temperature. Non-thermal
contributions come from photoelectric emission, photon
absorption/emission and $H_2$ molecule formation.
Let ${\cal R}$ be the ratio of the rates of increment of the
grain mean square angular momentum about the rotation axis $z$, $\langle
\Delta  J_z^2 \rangle$, due to this process with respect to atom collisions;
then
$${\cal R}={N_e \langle \Delta J_z^2 \rangle_e\over N_H \langle
\Delta  J_z^2 \rangle_H}={m_e\over m_H} {N_e\langle \epsilon\rangle_e\over
N_H\langle\epsilon \rangle_H }.\eqno(3.2)$$
In the previous formula $N_e$ and $N_H$ are the rates of photoelectric
ejection and of atom collisions per unit area, respectively; $\langle
\epsilon\rangle_e$
and $\langle\epsilon \rangle_H$ are the average kinetic energies corresponding
to the two processes. For negative grain charge,
$N_e=17.5{\cal F}_{\nu_L} Y_{\infty} \langle Q_{abs}(a)\rangle
\sim 8.75\times 10^4$ (cgs units);
$\langle \epsilon\rangle_e\sim
(\langle \epsilon\rangle-B)/2\sim 2$~eV, where $\langle \epsilon\rangle$  is
the spectrum average photon energy (Draine 1978);  it follows
${\cal R}_e\sim 5\times 10^{-3}$.
We have checked the importance of rotational excitation due to diffuse galactic
UV  photons and $H_2$ molecule formation. The first one is completely
negligible, while the second may very marginally contribute to
$T_{eff}$,
since ${\cal R}_{H_2}$, analogous to the photoelectric ${\cal R}$, is
${\cal R}_{H_2}= 0.025$. This low value can be understood observing
that the average
energy release following a $H_2$ molecule formation ($\sim 0.2$~eV) is
substantially smaller than $k T_e$. Thus
$$T_{eff}={1\over 2} T_e\left[1+\left({T_d\over T_e}\right) +\sum_j{\cal R}_j
\right] \simeq{1\over 2} T_e,\eqno(3.3)$$
since $T_d\ll T_e$; $j$ runs through all the above mentioned processes.
\par
Assuming spherical grains, upon substitution of eq. (3.3) into eq. (3.1),
we obtain
a simple relation between the radius and the rotational velocity
$$a= 5.94\times 10^{-4} T_{eff}^{1/5} \omega^{-\gamma}=3.26\times
10^{-3} \left({T_e\over 10^4{\rm K}}\right)^{1/5}=\beta \omega^{-\gamma},
(cgs units) having assumed a grain material density
$\delta=3.3$~g~cm$^{-3}$. According
to eqs.(3.2)-(3.4) we will use the value $\beta=3.26\times 10^{-3}$
(cgs units).
\par
Finally, since grains are charged and rotating, radiative damping due to dipole
emission may slow down the rotation rate. For a grain with an electric
dipole moment $d\simeq \xi Zea$ ($\xi\sim 0.01$)
the radiative damping time becomes equal to the collisional damping time
for
$$a_d=10^{-7}\left({\xi U\over V}\right)^{2/5}
\left({n_e \over 0.1{\rm cm}^{-3}}\right)^{-1/5}
\left({\delta \over 3.3{\rm g~cm}^{-3}}\right)^{-1/5}
\left({T_e \over 10^4{\rm K}}\right)^{1/10}~~{\rm cm},\eqno(3.5)$$
Hence, rotation of very small grains ($a\simlt 10$~\AA) will be
affected by radiative damping.
\par
It is appropriate at this point to pose the question about the possible
grain centrifugal disruption as well. Grains are destroyed if the centrifugal
stress $S_c \sim 3 I \omega^2/8\pi a^3 > \tau$: thus, it must be
$\omega < \sqrt{5\tau/\delta}~a^{-1}=1.23\times 10^5 a^{-1}$~Hz.
{}From a comparison with eq. (3.4), it appears that grains may easily
survive the centrifugal stress.
\par
There are several ways in which a grain may acquire dipole
moments: for instance, if the grain is not perfectly spherical,
it is rather possible that its center of mass and its center of
charge do not coincide. However, even for a spherical grain
a net electric dipole moment $d$ is likely to arise from
statistical fluctuations in the distribution of the charge
trapped in or on the grain. A spherical grain with electric potential
$U=Ze/a$ and capacity $C=a$, has a number of charges on each
hemisphere equal to $Z_{1/2}=UC/2e=Ua/2e$; the
statistical fluctuation of this number is roughly
$\sqrt{Ua/2e}$. Thus, since the length of the dipole is $\simeq
a$, the average dipole moment is $\langle d \rangle \simeq
\sqrt{Z/2} e a$. Note that, in this way,
the electric dipole is only a few percent
of the value that one would have obtained simply setting
$d=Zea$.
\par
The volume emissivity $\epsilon_\nu^d$ from the rotating, charged grains
can be easily calculated once the dust size distribution is specified.
We will assume for the latter a power law of the form
$${dn\over da}=A_i n_H a^{-q}~~~~{\rm for}~~a_0\le a\le a_1,\eqno(3.6)$$
where $n_H$ is the hydrogen density,
which reduces to the standard Mathis \etal (1977) (MRN) distribution if $q
\sim 3.5$;
in this case, the coefficient $A_i=3.37 \mu m_H \sigma/8\pi \delta \sqrt{a_1}$,
(where $\mu$ is the mean molecular weight and $\sigma\sim0.01$ is the
dust-to-gas ratio) as obtained normalizing the distribution with $\sigma$.
\par
Using eqs. (2.3), (3.4), (3.6) we write
$$\epsilon_\nu^d d\omega= {1\over 3} {e^2\eta\over c^3}
\gamma\beta^{4-q} A_i n_H \omega^{3-4\gamma+\gamma q} d\omega,
{}~~~~~~{\rm for}~
\left({a_1\over \beta }\right)^{-{1\over\gamma}} \le \omega \le \left({a_0\over
\beta }\right)^{-{1\over\gamma}}.\eqno(3.7)$$
Eq. (3.7) contains all the
physics introduced in the calculation so far; its implications are discussed
in the next \S.

\bigskip
\centerline{\sl 4. OBSERVATIONAL IMPLICATIONS}
\smallskip
In the previous \S~we have predicted the amount of radio emission expected
from dust immersed in a warm ($T_e\sim 10^4$~K), low-density ($n_e\sim
0.1$~cm$^{-3}$) gas, exposed to a diffuse UV spectrum. Here we evaluate the
possibility to detect the predicted emission.
\par
First we note that the wavelength range of the emission falls in a spectral
band that
can be easily observed. In fact, if we use $a_1=0.25\mu$m and $a_0=3$~\AA,
as suggested from the previous analysis, emission is expected  in the
frequency range ($\nu=\omega/2\pi$)
$$ 3.1\times 10^4 \le \nu \le 6.2  \times 10^{11}~~~{\rm Hz}.$$
In absence of radiative damping, small grains contribute to the 100
GHz band and the position
of the high frequency cutoff $\nu_c$ turns out to be a powerful tracer of the
smallest
sizes in the dust distribution; Fig. 3 (left panel) shows the dependence
of $\nu_c$ on $a_0$ for different temperatures.
In reality the upper limit will be decreased to lower frequencies
because of the radiative damping; in the present simple approach,
however, we have not investigated the modification of the spectrum due
to this effect.
Note that $\nu_c$ depends
very weakly on the model parameters (through ${\cal R}$) and also on
the temperature $T_e$. The knowledge of the
minimum grain size (and possibly a test of the MRN distribution)
would be a highly valuable piece of information on the dust properties.
\par
There is no doubt that the detection of the radio emission we predict would
be the major consequence of this paper; to evaluate this possibility
we compare the emissivity per H nucleus given by eq. (3.7) with the
free-free emissivity, $\epsilon_\nu^{ff}$
in the same frequency range by the ionized gas surrounding the dust.
We suppose that the medium is optically thin, which is certainly a good
approximation either for the gas and the dust at these frequencies;
then
$$\epsilon_\nu^{d}=1.36\times10^{-37} \left({\nu\over {\rm 100~GHz}}
\right)^{2.8}~~~ {\rm ergs~{\rm s}^{-1}~{\rm Hz}^{-1}};\eqno(4.1)$$
$$\epsilon_\nu^{ff}=6.78\times 10^{-41} \left({n_e\over 0.1{\rm cm}^{-3}
}\right) \left({T_e\over 10^{4}~{\rm K}}
\right)^{-1/2} g(\nu, T_e)~~~{\rm ergs}~{\rm s}^{-1}~{\rm Hz}^{-1},\eqno(4.2)$$
where $g(\nu,T_e)=(\sqrt{3}/\pi)\log(5\times 10^7 T^{3/2}/\nu)$.
The comparison between the two fluxes is shown in the right panel of
Fig. 3.
\par
The dust spectrum should present a steep positive spectral index $\alpha$
which depends only on the size distribution index $q$, since $\gamma$ is
fixed and determined by the microphysics; for a MRN distribution $\alpha=2.8$.
This should reduce the possibility of confusion with the flat free-free
spectrum and Fig. 3 clearly demonstrates that the emission can
in principle be detected.
\par
An intriguing hypothesis can be suggested about the observed vertical dust
filaments emerging from many spiral galaxy disks and immersed in the H$\alpha$
emitting gas (Sofue 1987; for a review see Dettmar 1992).
We propose that these filamentary shapes arise because a magnetic field ${\bf
B}$  aligned with the filament axis is present and the charged dust is tightly
coupled to the field. Such a vertical configuration of the field
could be produced by the enhancement of the Parker instability driven by
radiation pressure on charged grains (Ferrara 1993).
Thus,  the field lines act as tracks for
the grain motion and, viceversa, dust would represent an excellent tracer
of the ${\bf  B}$-field topology.
A necessary condition for this to be true is that the Larmor
radius, $r_L$, for the grains must be much smaller than the transverse size
of the filaments. Using the previous results, the maximum value of $r_L$ is
$$r_L= {4\pi \delta c\over 3 e \eta}{v\over B} a_1^2=0.12
\left({v\over 100 {\rm km~s}^{-1}}\right)\left({B\over \mu{\rm G}}\right)^{-1}
\left({a_1\over 250 {\rm nm}}\right)^2~{\rm pc}.\eqno(4.3)$$
Thus, even for models predicting high velocities of the grains due to
radiative acceleration (Ferrara \etal 1991), grains are strictly tied to
the magnetic lines. A definite test of this interpretation would be
represented by optical polarization measurements in which a correlation between
the degree of polarization due to aligned grains and extinction toward
background objects is found.
\bigskip
\centerline{\sl APPENDIX}
\smallskip
\centerline{\sl THE DETAILED BALANCE EQUATION}
The detailed balance equation we have used to determine the grain charge $Z$ is
$$n_p\left\{\left({8 k T_e\over \pi m_p}\right)^{1/2}\left[
{\tilde J}_{\nu_p,\tau_p} (a, Z, T_e)s_p + \delta_p(a, Z, T_e)\right]+
{\tilde J}_{pe}(a, {\cal F}_\nu)\right\}=$$
$$n_e \left\{\left({8 k T_e\over \pi m_e}\right)^{1/2}
\left[ {\tilde J}_{\nu_e,\tau_e} (a, Z, T_e)s_e - \delta_e(a, Z, T_e)
\right]\right\},\eqno(A1)$$
where the index $p$,$e$ stands for protons and electrons, respectively;
${\tilde J}_{\nu,\tau}$  are the collisional charging rates;
$\delta$  are the secondary emission rates; ${\tilde J}_{pe}$ is the
photoelectric charging rate; $s$ is the sticking probability.
${\tilde J}_{\nu,\tau}$ are taken from Draine \& Sutin (1987);
${\tilde J}_{pe}$ is taken from Draine (1978); $\delta$ and $s$ are
taken from Draine \& Salpeter (1979).
Eq. (A1) must be solved together with the field emission condition (Draine
\& Sutin 1987) that limits the value of the charge:
$$-1-0.7\left({a\over{\rm nm}}\right)^2< Z < 1 + 21\left({a\over{\rm
nm}}\right)^2. \eqno(A2)$$
\par
When only ion and electron collisions are considered in a pure H plasma,
eq. (A1) can be written as
$$\left({m_e\over m_p}\right)^{1/2}={{\tilde J}_{\nu_e,\tau_e} (a, Z, T_e)\over
{\tilde J}_{\nu_p,\tau_p} (a, Z, T_e)},\eqno(A3)$$
with $s_e=s_p=1$.
For $\vert\nu\vert\ll 1$, ${\tilde J}\to (1-\nu/\tau)$ for $\nu < 0$
and ${\tilde J}\to e^{-\nu/\tau}$ for $\nu > 0$, where $\nu=Ze/q$,
$\tau=akT/q^2$
(see Draine \& Sutin 1984) and $q$ is the charge of the colliding
particle. Thus
$$\left({m_e\over m_p}\right)^{1/2}={e^{\psi}\over 1-\psi},\eqno(A4)$$
where $\psi=Z/\tau$. The solution of eq. (A4) is $\psi=-2.504$ and
eq. (2.3) follows accordingly.

\bigskip
\centerline{\sl ACKNOWLEDGMENTS}
\smallskip
We are deeply indebted to Dr. P. G. Martin and Dr. F. Rouleau for having
made available their data on the grain dielectrical constants in advance
of publication; we thank C. Heiles for useful discussions, and an
anonymous referee for pointing out an error in a previous version and
for stimulating remarks.

\bigskip
\parindent=0pc
\parskip=6pt
\centerline{\sl REFERENCES}
\vskip 2pc
\ref Chandrasekhar, S. 1943, Rev. Mod. Phys., 15, 1

\ref Dettmar, R.-J. 1992, Fundamental of Cosmic Physics, 15, 143

\ref Draine, B. T. 1978, ApJSS, 36, 595

\ref Draine, B. T.  \& Salpeter, E. E. 1979, ApJ, 231, 77

\ref Draine, B. T.  \& Lee, H. M. 1984, ApJ, 285, 89

\ref Draine, B. T.  \& Sutin,  1984, ApJ, 320, 803

\ref Erickson, W. C. 1957, ApJ, 126, 480

\ref Ferrara, A, Ferrini, F. Franco, J. \& Barsella, B. 1991, ApJ, 381, 137

\ref Ferrara, A. 1993, in preparation

\ref Franco, J. J., Ferrini, F., Ferrara, A., \& Barsella, B. 1991, ApJ, 381,
137

\ref Hoyle, F. \& Wickramasinghe, N. C. 1970, Nature, 227, 473

\ref Huffman, D. R. 1975, ApJS, 34, 175

\ref Keppel, J. W., Dettmar, R.-J., Gallagher, J. S., \& Roberts, M. S. 1991,
ApJ, 374, 507

\ref Martin, P. G. \& Rouleau, F. 1991, Extreme UV Astronomy, eds. S. Malina
     \& S. Bowyer, (Oxford:Pergamon), 341

\ref Mathis, J. S. 1990, ARAA, 28, 37

\ref Mathis, J. S., Rumpl, W. \& Nordsiek, K. H. 1977, ApJ, 217, 425 (MRN)

\ref Purcell, E. M. 1979, ApJ, 231, 404

\ref Reynolds, R. J. 1990, The Galactic and Extragalactic Background Radiation,
     eds. S. Bowyer \& C. Leinert, (Dordrecht:Kluwer), 165

\ref Reynolds, R. J. 1991, The Interstellar Disk-Halo Connection,
     ed. H. Bloemen, (Dodrecht:Kluwer), 67

\ref Reynolds, R. J. 1993, 3rd Annual Maryland Meeting, Back to the Galaxy,
     eds. S. S. Holt and F. Verter, (Chicago:AIP), 156

\ref Sofue, Y. 1987, PASJ 39, 547

\ref Spitzer, L. 1941, ApJ, 93, 369

\vfill\eject

\bigskip
\centerline{\sl FIGURE CAPTIONS}
\smallskip
\vskip0.2in
{\bf Figure 1} $Q_{abs}(a,\epsilon)$ for spherical
silicate grains using the  dielectric constants given by Martin \& Rouleau;
numbers indicate the grain radius in nm.

\vskip0.1in
{\bf Figure 2}
Numerical solutions of the detailed balance equation (see Appendix).
{\bf Left:} grain charge as a function of $a$ for different values of
${\cal F}_{\nu_L}$, as shown by numbers; {\bf right:} rates for
the various charging processes for the standard case ${\cal F}_{\nu_L}=10^4$;
{\it long-dashed line-} electron collisions;
{\it solid-} proton collisions;
{\it short-dashed -} photoelectric effect;
{\it dotted-} proton secondary emissions;
{\it dot-dashed-} electron secondary emission.

\vskip0.1in
{\bf Figure 3} {\bf Left:} dependence of the cutoff frequency $\nu_c$ on
the minimum size of the grain distribution, $a_0$, for different temperatures:
{\it solid-} $T_e=10^4$~K;
{\it dotted-} $T_e=7000$~K;
{\it dashed-} $T_e=1.2\times 10^4$~K.
{\bf Right:} comparison between the free-free
emissivity ({\it flat dotted line}) from a gas with $T_e=10^4$~K
and $n_e=0.1$~cm$^{-3}$
and dust emissivity ({\it steep dotted line}) per H nucleus;  the {\it solid}
line is the sum of the two. The dashed region denotes the range of
frequency in which the spectrum may be modified by radiative damping
(see text).
In the upper right corner the cutoff frequency  $\nu_c$ is shown.

\vfill\eject
\bye